\documentclass[11pt]{article}

\usepackage[legalpaper, margin=2.3cm]{geometry}
\usepackage{authblk}
\usepackage{graphicx}
\usepackage{amsmath}
\usepackage[style=nature]{biblatex}
\title{\textbf{High-Quality Pulse Compression Using a Hybrid All-Bulk Multipass Cell Scheme}}
\author[1,*]{V. W. Segundo Staels}
\author[1,2]{E. Conejero Jarque}
\author[1,2]{J. San Roman}
\affil[1]{Grupo de Investigación en Aplicaciones del Láser y
Fotónica, Departamento de Física Aplicada, Universidad de
Salamanca, E-37008 Salamanca, Spain}
\affil[2]{Unidad de
Excelencia en Luz y Materia Estructuradas (LUMES),
Universidad de Salamanca, Salamanca 37008, Spain}
\affil[*]{vwsstaels@usal.es}

\begin{document}

\maketitle

\begin{center}
\begin{minipage}{0.9\textwidth}
    \small 
    \textbf{Abstract}  We present a detailed numerical study of ultrashort pulse compression using a three-stage hybrid all-bulk multipass cell scheme. By operating in the enhanced frequency chirp regime, we achieve the compression of pulses from around 180 fs to 4 fs pulse duration (a  total compression factor above 45), with side lobes contributing with intensity values lower than 0.2 \% of the peak intensity. Optimal conditions for the enhanced frequency chirp regime propagation have been identified, enabling smooth spectral broadening and high-quality temporal profiles. The first two stages are based on bulk multipass cells to achieve a controlled spectral broadening, while the third stage consists of a thin plate to reach the spectral broadening needed for few cycle pulses without leaving the enhanced frequency chirp regime.     
\end{minipage}
\end{center}

\section{Introduction}

Ultrashort, high-intensity laser pulses have revolutionized various scientific and industrial fields, driving relevant advances in areas such as strong-field physics, attosecond science, multiphoton microscopy, and materials science \cite{KrauszMisha,gamaly_physics_2011,nolte_ultrashort_2016}. These applications often demand pulses with durations of a few tens of femtoseconds or even shorter, enabling the investigation of ultrafast processes with unprecedented temporal resolution. To overcome the bandwidth limitations of conventional lasers and achieve the desired shorter pulse durations, various post-compression techniques have been developed. Most of these techniques rely on the principle of spectrally broadening the laser pulse through the introduction of a nonlinear phase modulation and then compensating the acquired spectral phase using an external compression device, such as a diffraction grating or chirped mirrors \cite{nagy_high-energy_2021,khazanov_post-compression_2022}.

Among the various post-compression methods, nonlinear propagation in multipass cells (MPCs) have emerged as a promising technique, positioning at the forefront of these methods for high average power laser systems \cite{schulte_nonlinear_2016,hanna_nonlinear_2021,viotti_multi-pass_2022}. MPCs are based on the propagation of a laser pulse through a nonlinear medium contained in a resonant optical cavity, such as a Herriott cell \cite{herriott_off-axis_1964}. The cavity is designed so that the pulse makes multiple passes through the medium, accumulating a significant nonlinear phase shift through self-phase modulation (SPM).

Advantages of MPCs include high transmission efficiency, excellent beam quality, scalability to high power, and flexibility in the choice of nonlinear medium. It has also been verified that it is possible to achieve self-compression of laser pulses in setups based on multi-pass cells \cite{jargot_self-compression_2018,grobmeyer_self-compression_2020,carlson_nonlinear_2022}. In particular, gas-filled MPCs offer smooth nonlinear response, broad transparency range, and reduced thermal effects, but are limited by their lower nonlinearity and more complex setup. Multipass cells including bulk media, on the other hand, show higher nonlinear responses and simpler setups. While they may present challenges, such as spectral bandwidth limitations and thermal effects arising from heat accumulation in the nonlinear medium under high average powers \cite{seidel_factor_2022,mei_thermodynamic_2022,omar_hybrid_2024}, these issues can often be mitigated by carefully tuning the input parameters to operate in a favorable regime. Recent works have demonstrated that bulk MPCs can achieve efficient few-cycle pulse compression with remarkable stability and spectral broadening, showing their potential as a compact and effective post-compression scheme \cite{liu_few-cycle_2024,fritsch_all-solid-state_2018}. Another key benefit of bulk MPCs, compared to propagation through thin plates alone, is their ability to better preserve the spatio-spectral homogeneity of the pulse \cite{seidel_factor_2022}. 

A possible problem when shortening pulses with large compression factors is the loss of temporal quality \cite{escoto_temporal_2022}. Improving the pulse temporal quality in MPC post-compression can be achieved using various advanced techniques, such as employing higher-order dispersion management or dividing the compression into multiple stages \cite{escoto_temporal_2022}, by filtering the self-phase-modulated generated part of the spectrum \cite{buldt_temporal_2017}, or via the nonlinear elliptical polarization rotation \cite{pajer_high_2021,song_temporal_2022,kaur_simultaneous_2024,escoto_improved_2024}. A different approach is to exploit the enhanced frequency chirp regime (EFCR), which involves working in a region of parameters in which the SPM and the group velocity dispersion are both relevant effects, generating smooth spectra and, therefore, helping to enhance the temporal quality of the compressed pulse.

Although the EFCR was first reported in the 1980s in the context of optical fibers \cite{Grischkowsky:82,tomlinson_compression_1984}, it has been recently extended to gas-filled MPCs. Theoretical studies \cite{benner_concept_2023,staels_numerical_2023} and experimental demonstrations \cite{karst_dispersion_2023} have confirmed the effectiveness of EFCR in MPCs to achieve short pulses with improved temporal quality and increased peak pulse power by using dispersive mirrors or tuning the gas pressure and input pulse energy.
This work extends the theoretical study of MPCs operating in the EFCR to bulk materials, showing the possibility of obtaining clean ultrashort pulses with minimal secondary structure close to the single-cycle regime in a robust all-bulk configuration.

\section{Methodology and MPC design}
\subsection{Numerical model}

To investigate the generation of clean ultrashort pulses in bulk MPC setups, we simulate the nonlinear propagation of the electric field envelope $A(x, y, t, z)$ using a (3+1)D split-step Fourier method, which solves the propagation equation by alternating between linear and nonlinear steps. The linear part, which includes diffraction and material dispersion, is computed in the spectral domain. Specifically, the evolution of the field is given by:
\begin{equation}
\tilde{A}(\omega, k_x, k_y, z + dz) = \tilde{A}(\omega, k_x, k_y, z)\, \exp\left[i\, dz\, k(\omega) \sqrt{1 - \frac{k_x^2 + k_y^2}{k(\omega)^2}}\right],
\end{equation}
where $dz$ is the propagation step and $k(\omega)$ the wavenumber. The nonlinear part is calculated in the temporal domain and accounts for the optical Kerr effect (including self-phase modulation and self-focusing), self-steepening, and stimulated Raman scattering. This step is expressed as:
\begin{equation}
A(t, x, y, z + dz) = A(t, x, y, z)\, \exp\left[i\, k_0\, dz\, n_2\, \Psi(A)\right],
\end{equation}
where $n_2$ is the nonlinear refractive index and the nonlinear response term is defined by:
\begin{equation}
\Psi(A) = \hat{T} \int R(t')\, |A(t - t')|^2\, dt'
\end{equation}
Here, $\hat{T}$ is the operator associated with the self-steepening effect, and $R(t')$ represents the nonlinear response function, including both instantaneous (Kerr) and delayed (Raman) contributions. A more detailed description of the numerical model can be found in \cite{staels_numerical_2023}.

\subsection{Multipass cell design: conditions to be in the EFCR}
There are several conditions that have to be fulfilled in order to be in the EFCR that will help us to define a good MPC setup. First, both the self-phase modulation (SPM) and the material dispersion must be relevant effects during the propagation of the pulse. One way to ensure this condition is to maintain the interaction length ($L_I$) between the nonlinear and the dispersion lengths: ($L_{NL}<L_I<L_D$) \cite{staels_numerical_2023}, where the nonlinear and dispersion lengths are defined as: $L_{NL}=2/(k_0 n_2 I_0)$ and $L_D=T_0^2/|\beta_2|$ \cite{Agrawal}.
In these definitions $k_0=\omega_0/c$ is the wavenumber at the central wavelength of the input pulse, $n_2$ is the nonlinear refractive index of the nonlinear medium, $I_0$ is the input peak intensity, $T_0$ is the temporal duration of the input pulse and $\beta_2$ is the group velocity dispersion (GVD) of the nonlinear medium. We have used the full width at half maximum (FWHM) of the intensity temporal profile as $T_0$ to calculate the dispersion length. Second, to ensure stable propagation and avoid the appearance of any self-focusing dynamics, we impose that the bulk material thickness ($L$) must be much shorter than the collapse length ($L_C$), specifically we propose $L<L_C/10$. The collapse length, defined following Marburger's formula \cite{marburger_self-focusing_1975}, estimates the distance over which a high-peak-power beam would theoretically collapse due to self-focusing in the absence of ionization, and is given by: $L_C =0.367 L_{DF}/ \sqrt{[(P_{in}/P_{cr})^{1/2}-0.852]^2-0.0219}$, where $L_{DF}=k_0 w_0^2/2$ is the diffraction length, $w_0$ the spatial width of the input beam, and $P_{in}$ and $P_{cr}$ correspond to the input and critical peak powers, respectively. 
Since in our case $P_{in} \gg P_{cr}$, limiting the material thickness well below $L_C$ prevents the development of strong nonlinear effects such as self-focusing and ionization, thus preserving the beam quality.
Notice that these conditions are scalable and represent a quite large family of different experimental configurations showing the same general dynamics (see the Supplemental material for a more detailed discussion on this scaling).
Last, if we want to achieve ultrashort pulses within the few-cycle regime, as the spectral broadening is limited due to the important stretching of the pulse in the EFCR, we will need to build a cascade setup, based on several MPC stages, paying special attention to the smoothness of the spectral structure of the pulse at the entrance of each stage to avoid any coupling between the spectral modulations into the propagation dynamics, which would deteriorate the pulse cleanness.

As a representative case, we start with a standard 800 nm Gaussian beam with pulse duration $T_0=177$ fs (FWHM) ($ t_p=150$ fs) and with $220~\mathrm{\mu J}$ of energy. We use two identical fused silica thin plates located on the mirrors of the MPC, while keeping the MPC in vacuum. Regarding the cavity, we use two concave mirrors, both with 7.3 m radius of curvature, separated by 40 cm, so that the linear fundamental mode has a waist of $500~\mathrm{\mu m}$ and is located at the center of the cavity. 
Although a tighter focusing configuration (near-confocal) would allow the use of thicker plates, it would require much larger spatial grids, making simulations computationally prohibitive. That is why in this work we chose a looser focusing setup that eases numerical modeling while preserving the general strategy.
We assume a perfect coupling of the beam into the fundamental mode of the cavity and we neglect reflection and transmission losses at the surfaces. It is worth noting that in a real system these losses would impact both the efficiency and the resulting spectrum during the initial round trips. Losses occurring in the later part of the propagation, after spectral broadening has saturated, would primarily affect the output energy. Therefore, we expect that the overall efficiency of the setup would be the property most affected by such losses. These losses could be partially compensated tweaking the initial pulse parameters to obtain the desired output pulse. 

\begin{figure}[htbp]
    \centering
    \includegraphics[width=0.45\linewidth]{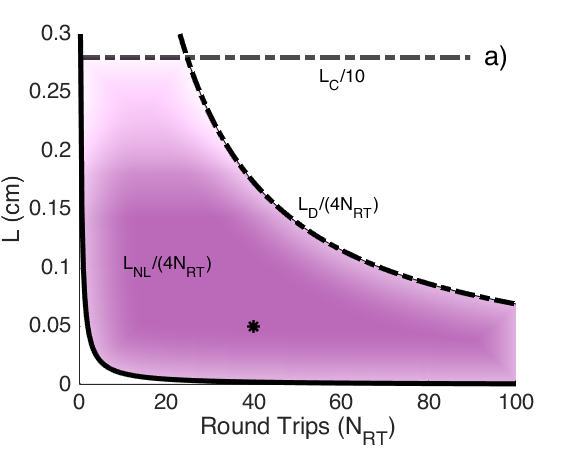}
    \includegraphics[trim={-1cm 0 0 0},width=0.47\linewidth]{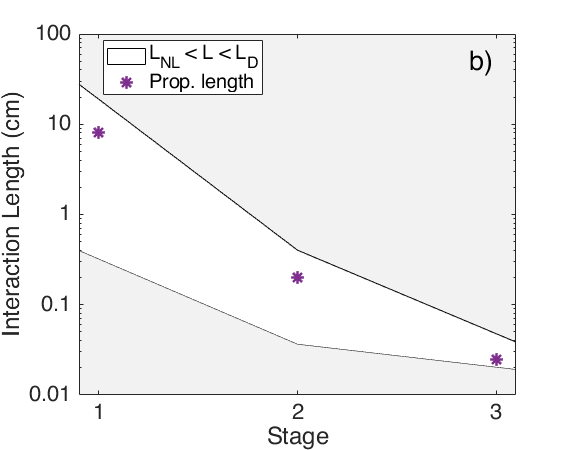}
    \caption{Width of the fused silica plates as a function of the number of round trips (left panel) for the first stage. The lilac region represents the (L,$\text{N}_{\text{RT}}$) region for which the propagation is in the EFCR. The right panel shows the interaction length in each stage to operate in the EFCR.}
    \label{fig:design}
\end{figure}

To decide the width of the fused silica plates ($L$) and the number of round trips ($N_{RT}$), we have to consider the conditions to be in the desired robust EFCR. The input peak intensity at the plates is $2.86\times10^{11}~\mathrm{W/cm}^2$ with a corresponding peak power of $1.7$ GW. While the peak power is much greater than the critical power of fused silica at 800 nm, the peak intensity remains below the $7.4~\mathrm{TW/cm}^2$ limit, where ionization can be neglected \cite{toth_single_2023}. Taking into account the input peak intensity and the temporal duration of the pulse, and that $\beta_2=36.163~\mathrm{fs^2/mm}$ (obtained from the Sellmeier's formula \cite{malitson_interspecimen_1965}) and $n_2=2.22\times10^{-20}~\mathrm{m^2/W}$ for fused silica at 800 nm \cite{schiek_nonlinear_2023}, we obtain the following values for the nonlinear, dispersion and collapse lengths: $L_{NL}=0.4$ cm, $L_D=27.7$ cm and $L_C=2.8$ cm. As the beam goes through the fused silica plates four times per round trip, the interaction length for our configuration can be written in terms of the plate width and the number of round trips as $L_I=4N_{RT}L$, so that the two first conditions are described by: $L_{NL}/(4N_{RT})<L<L_D/(4N_{RT})$ and $L<L_C/10$. The lilac filled area of Fig. \ref{fig:design}(a) shows the possible widths of the fused silica plates to fulfill these two conditions, depending on the number of round trips of the design. For this first stage, we have chosen a standard number of round trips, 40, so according to the limitations of the dispersion, nonlinear, and collapse lengths, a pair of fused silica plates of $500~\mathrm{\mu m}$ perfectly fulfill all the conditions.  We have represented our choice with an asterisk in Fig. \ref{fig:design}(a). This selection is intended to illustrate a specific near-optimal working case, nevertheless, other nearby configurations are equally valid and lead to comparable results, as shown in the Supplementary Material. 

A similar procedure will be performed for the next stages, but using the new input peak intensity achieved after the compression of the pulse obtained in the previous stage. Intuitively, the following MPC stages will start with shorter and more intense pulses, so they will have to use less material and/or less round trips to assure keeping the compression process in the EFCR. In our particular case of a cascade scheme with three stages, the last one becomes a single pass through a single fused silica plate instead of a MPC setup (see Fig. \ref{fig:design}(b) for the total nonlinear material width for the three stages proposed in this work). This is what we have called an all-bulk hybrid cascade MPC scheme, as not all the stages of the cascade scheme are MPC setups.

\section{Single stage compression results}

\begin{figure}[htbp]
    \centering
    \includegraphics[width=0.43\linewidth]{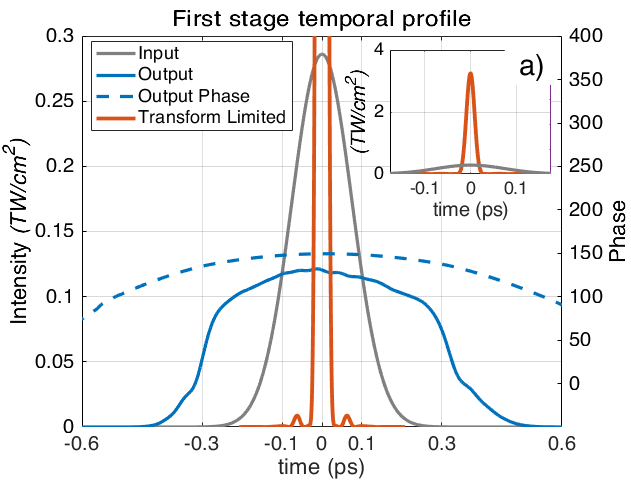}
    \includegraphics[trim={0 0 0 2cm},width=0.44\linewidth]{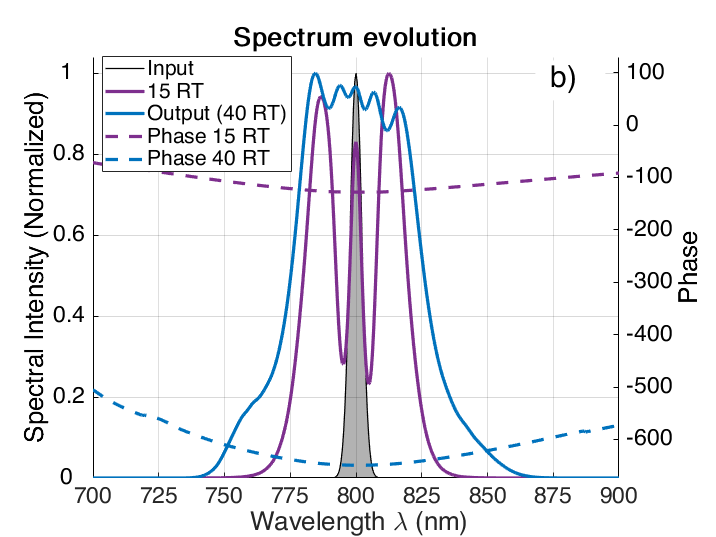}
    \includegraphics[trim={1cm 0 2cm 0},width=0.82\linewidth]{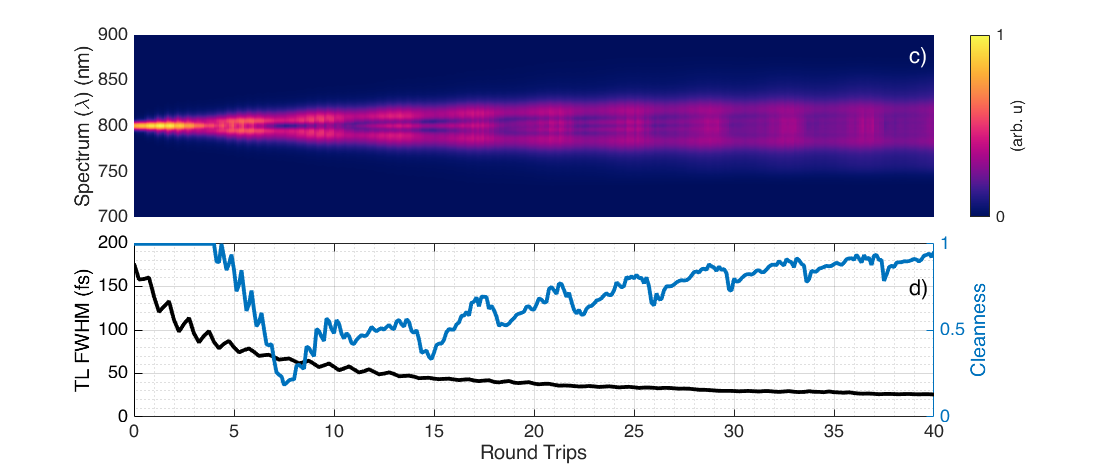}
    \caption{(a) Temporal on-axis intensity distribution of the input (gray), output (blue) and the corresponding Transform Limited (TL) of the output (orange) beam. The dashed-blue line shows the on-axis temporal phase of the output beam. The inset depicts the input and the TL output pulse beams in a different intensity scale to show the full height of the latter. (b) On-axis spectral distribution during propagation at the starting point (gray), after 15 (purple) and 40 (blue) round trips. The dashed-purple and dashed-blue lines show the on-axis spectral phase of the beam after 15 and 40 round trips, respectively. On-axis spectral intensity (c), and cleanness (blue) and on-axis TL pulse duration (black) (d) during the propagation in the MPC.}
    \label{fig:1stStage}
\end{figure}

With the configuration described above for the first stage (two $500~\mathrm{\mu m}$ fused silica glass plates located on the mirrors and 40 round trips) we obtain a clean short Transform Limited pulse (TL) at the end of the first stage, with a compression factor above 8 (from 177 fs FWHM to 20.7 fs), as shown in the inset box of Fig. \ref{fig:1stStage}(a). In Fig. \ref{fig:1stStage}(c) we show the spectral evolution that makes this possible. During the first round trips, the pulse undergoes significant spectral broadening due to the self-phase modulation induced by the high peak intensity present during this part of the propagation. The created spectrum is highly modulated because the dispersion accumulated is not enough yet to rearrange it, as can be seen in the purple curve in Fig. \ref{fig:1stStage}(b) which shows the on-axis spectral intensity obtained after 15 round trips. From the temporal point of view, the pulse stretches in accordance with the positive linear and nonlinear GVD and the nonlinear spatial readjustments. Obviously, there is an important decrease of the duration of the TL pulse (see Fig. \ref{fig:1stStage}(d)) although, during this first part of the propagation, it still shows considerable high side structures linked to the noticeable modulations of the corresponding spectrum. A quantifiable parameter to measure how well the spectral broadening behaves, in terms of pulse quality, is the spectral cleanness, closely related to the TL pulse structure, which gives us an idea about how well filled the spectrum is. We define the spectral cleanness $(\mathrm{SC})$ as the visibility of the spectral modulations, $\mathrm{SC}=2I_m/(I_M+I_m)$, where $I_M$ and $I_m$ are the highest and lowest spatially integrated spectral intensity values inside the spectral width \cite{staels_numerical_2023}. In general, the higher the $\mathrm{SC}$, the lower the side lobes of the TL pulse structure. In our case we begin with a perfect Gaussian pulse ($\mathrm{SC}=1$, without side lobes) and during this first part of the propagation the cleanness decreases in accordance to the modulated spectral broadening process (see Fig. \ref{fig:1stStage}(d)).  

After approximately 25 round trips, the spectral broadening stops increasing but the nonlinearity, together with the linear dispersion, begins to effectively fill the spectral modulations and broaden the spectral tails, two clear signals of the EFCR in which the propagation is taking place. We can also observe in Fig. \ref{fig:1stStage}(d) that the spectral cleanness keeps increasing, reaching values close to 1 again, which means that the side lobes of the TL pulse are disappearing. At the end, after 40 round trips, the maximum intensity of the first side lobe is below $0.3\%$ of the peak intensity, as shown in Fig. \ref{fig:1stStage}(a). To finish this discussion on the results for the first stage, we have examined if it could be beneficial to maintain the propagation of the beam a few more round trips in the cell and, furthermore, a more detailed description of the spectral phase acquired during the nonlinear propagation. We have checked that if we let the pulse propagate during a few more round trips, there is little impact in spectral broadening, the duration of the TL pulse or the final pulse cleanness. The main reason for the lack of significant improvement is the long temporal duration of the pulse after the first 40 round trips that inhibits any further nonlinear dynamics.

\begin{figure}[htbp]
    \centering
    \includegraphics[width=0.44\linewidth]{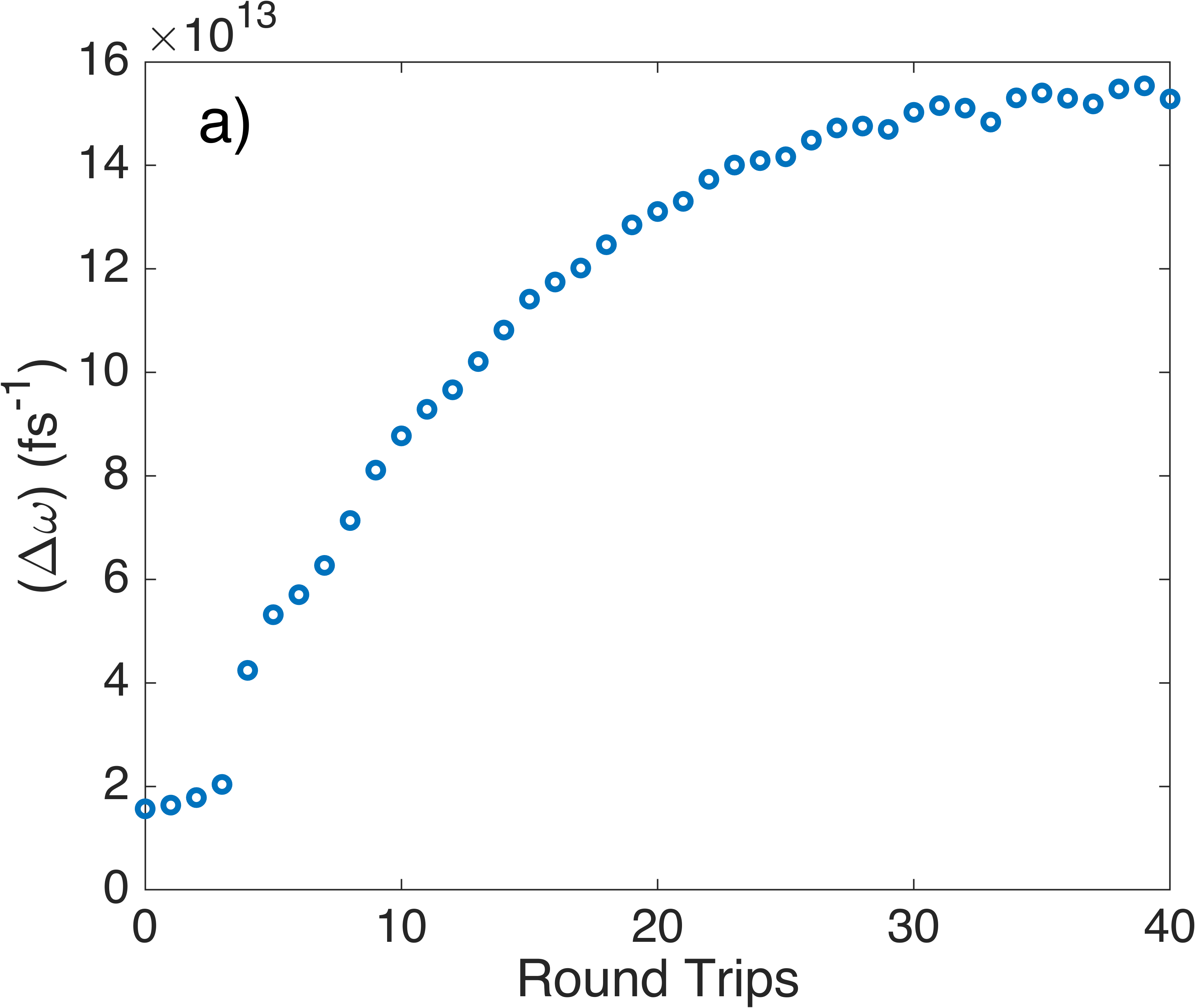}
    \includegraphics[width=0.44\linewidth]{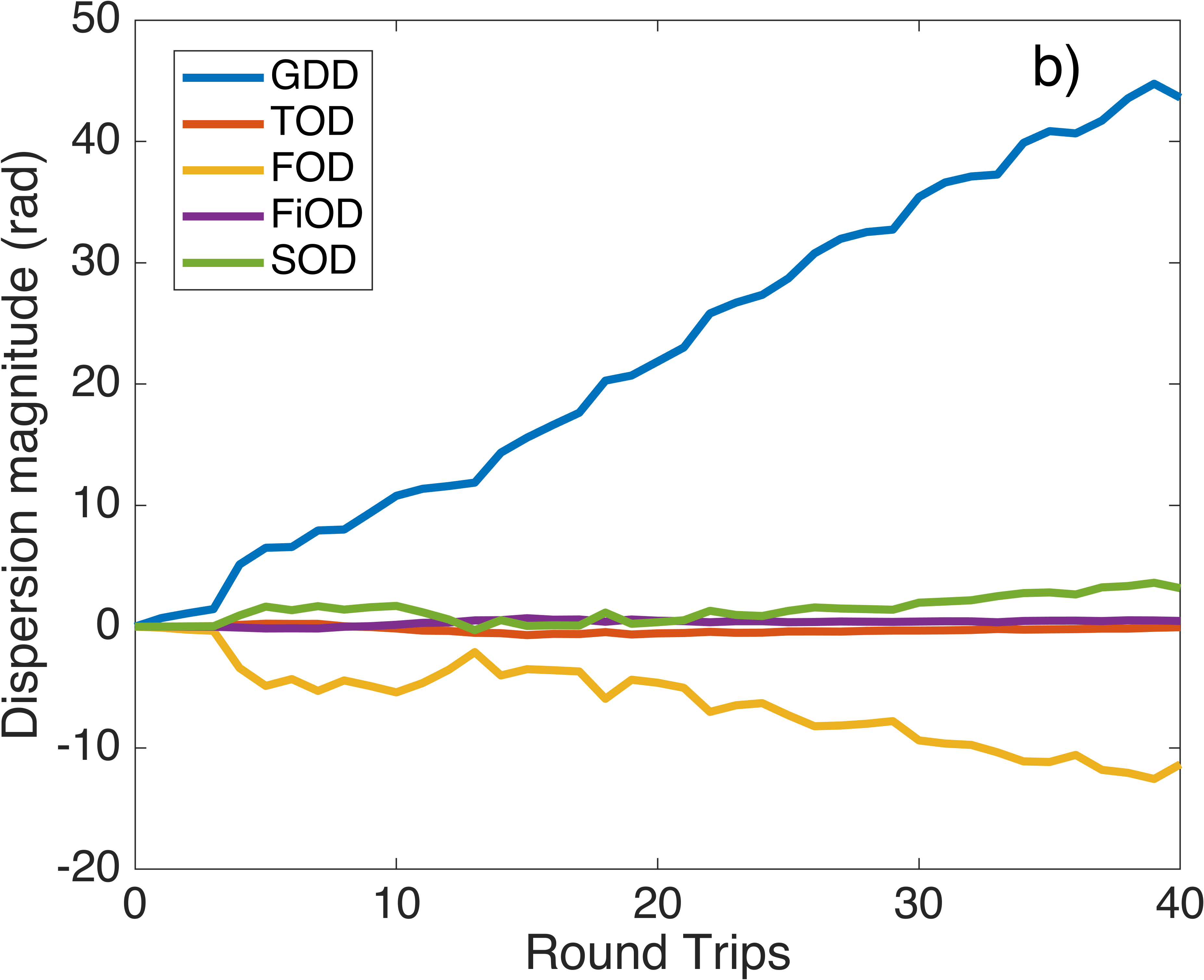}
    \caption{Evolution of the on-axis spectral width ($\Delta\omega$) (a) and the different contributions of the spectral phase up to sixth order (b) during the propagation. The phase contributions are presented as  GDD$(\Delta\omega)^2/2!$, the quadratic contribution, TOD$(\Delta\omega)^3/3!$ the cubic, etcetera.}
    \label{fig:DispEvo}
\end{figure}

Regarding the spectral phase of the pulse, we have analyzed the evolution of the first dispersion orders during the propagation: group delay dispersion (GDD), third order dispersion (TOD), fourth order dispersion (FOD), fifth order dispersion (FiOD) and sixth order dispersion (SOD). These values are calculated by fitting the temporal profile of the pulse at each propagation distance to the pulse obtained when adding a particular expansion of the spectral phase (up to sixth order) to the modulus of the corresponding spectrum. In Fig. \ref{fig:DispEvo}, we show the evolution of the on-axis spectral width, $\Delta\omega$ (a), and the different contributions of the spectral phase over the propagation, namely GDD$(\Delta\omega)^2/2!$, TOD$(\Delta\omega)^3/3!$, FOD$(\Delta\omega)^4/4!$, etcetera, (b), which provides insight about which terms contribute more. We can see that the main contribution comes from the quadratic term, the GDD, so compensating for this term will be enough to retrieve a short pulse. Nevertheless, the fourth order is not negligible and must also be taken into consideration if we want to retrieve clean pulses. 
Remarkably, odd-order terms play a much less significant role. This behavior can be understood from the fact that for largely chirped pulses, the accumulated spectral phase can be well approximated by the nonlinear temporal phase, which in our simulations remains sufficiently smooth and symmetric throughout propagation, thus avoiding the generation of strong odd-order contributions to the spectral phase \cite{Galvanauskas_fermann_ultrafast_2002}. Any temporal or spectral irregularities would lead to irregular phase modulations in the pulse, complicating compression. Therefore, maintaining a smooth temporal pulse profile is essential to facilitate clean compression.
Here, the most dominant terms are clearly the GDD and FOD. 

Finally, as we have a relatively complex spatio-spectral nonlinear evolution of the beam, we should verify that we are always below the damage threshold of the plates. For the case of a fused silica plate, the damage threshold of a 150 fs laser pulse at 800 nm is around $3~\mathrm{J/cm^2}$ \cite{chimier_damage_2011}, and we have checked that we are always well below that value, reaching a maximum fluence value of 0.11 (0.15 and 0.09) J/cm$^2$ at the first (second and third) stage. Additionally, the spatial homogeneity is well preserved throughout the propagation, as shown in the Supplement, with the average V-factor at the output calculated to be $99.98\%$, indicating an excellent degree of spatial uniformity in the output beam.
\section{Multi-stage compression scheme results}
\subsection{Second compression stage}

In order to further compress our pulse into a shorter one, we need another compression stage. We propose to use the compressed output pulse from the previous stage as the input pulse for a second MPC stage. This new stage will have the same parameters as the first one, except for the fused silica plate thickness and the number of round trips, which will be recalculated to ensure that the propagation remains in the EFCR. To do so, we need to know the peak intensity and the temporal FWHM duration of the input beam, which depends on the compression of the output pulse after the first stage. In Table \ref{table:1} we discuss three different situations based on using the Transform Limited pulse (TL) obtained from the first stage (top row), i.e. a perfect compression, or the pulse obtained after compensating up to second order (GDD) (middle row) and fourth order (GDD, TOD and FOD) (bottom row). With these values we choose the width of the fused silica plates and the number of round trips so that $L_{NL}/(4N_{RT})<L<L_D/(4N_{RT})$ and $L<L_C/10$ are fulfilled. As expected, the width of the plates and/or the number of round trips have to decrease notably. Our choice is to use two plates of 100 $\mu m$ width and 5 round trips, decreasing the $L_I$ from $8$ cm in the first stage to $0.2$ cm in the second, as shown in Fig. \ref{fig:design}(b).
\begin{table}[h!]
\centering
\begin{tabular}{ |c|c|c|c|c|c| } 
\hline
 & I ($\mathrm{W/cm}^2$) & FWHM (fs) & $L_D$ (cm) & $L_{NL}$ (cm)  &  $L_C$ (cm) \\
\hline
Fourier Limit & $3.25\times10^{12}$ & $20.7$ & $0.38$ & $0.04$ & $0.91$ \\ 
\hline
GDD & $2.25\times10^{12}$ & $26.0$ & $0.60$ & $0.05$ & $1.11$  \\
\hline
Up to FOD & $3.12\times10^{12}$ & $21.3$ & $0.40$ & $0.04$ & $0.93$ \\ 
\hline
\end{tabular}
\caption{Input pulse peak intensity and pulse duration at the entrance of the second stage after compensating the phase of the output pulse at the first stage in three different ways, and the corresponding dispersion, nonlinear and collapse lengths in each case.}
\label{table:1}
\end{table}

\begin{figure}[htbp]
    \includegraphics[width=0.44\linewidth]{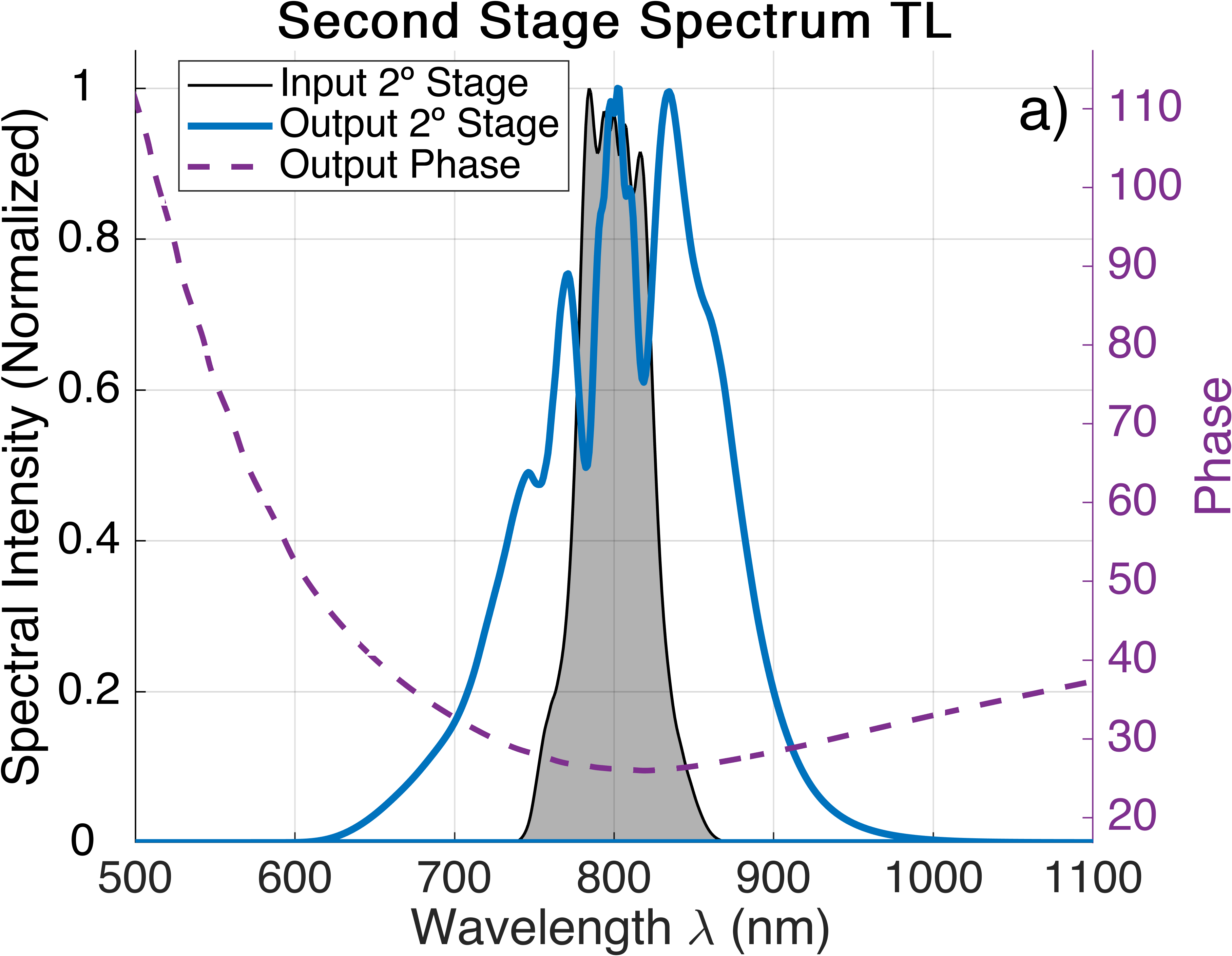}
    \includegraphics[width=0.44\linewidth]{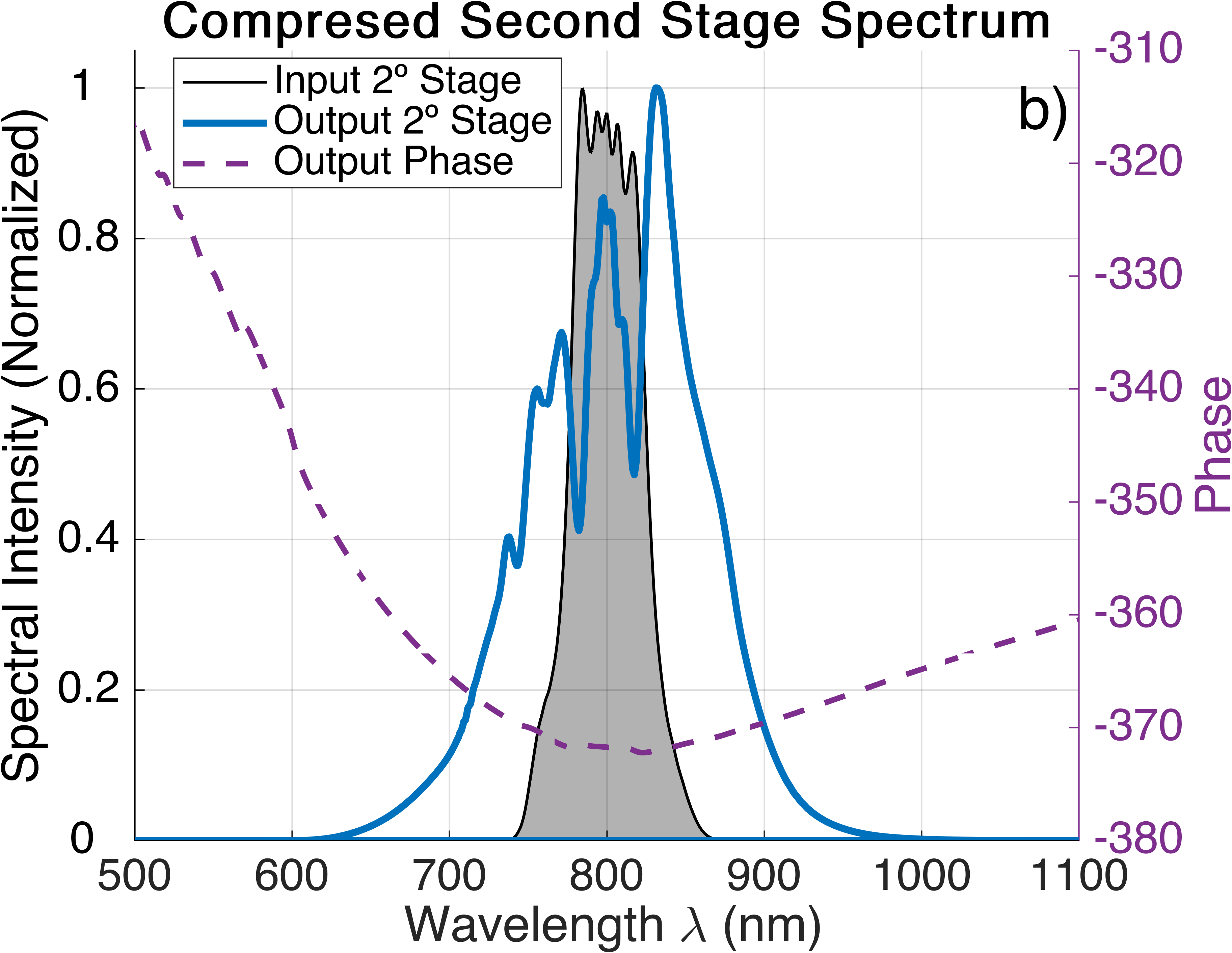}
    \centering
    \caption{ On-axis output spectra and phase obtained from the second stage when introducing (a) the TL pulse or (b) the pulse after compensating for GDD, TOD and FOD.}
    \label{fig:Demo3}
\end{figure}

In Fig. \ref{fig:Demo3}, we show the on-axis intensity and phase of the output spectrum after propagating 5 round trips in the second stage using the TL pulse (a) or the pulse after compensating the first three dispersion terms (b) as the input pulse. The phase added to the pulse in the non-perfect compression case contained -3515 fs$^2$ of GDD, -800 fs$^3$ of TOD and $238.6\cdot10^3$ fs$^4$ of FOD, which can be imprinted with custom compressors, for example, using chirped mirrors and/or grism-based systems \cite{dou_dispersion_2010}. The propagation of any of these two pulses induces dynamics akin to those observed in the first stage, allowing for further compression of the pulse. For the TL pulse case, Fig. \ref{fig:Demo3}(a), the spectral broadening is significant, with a relatively clean spectral phase and with a relatively high spectral cleanness $(\mathrm{SC}=0.64)$. This spectrum is compatible with an output TL pulse of $6.73$ fs FWHM duration and in which the $\mathrm{SC}$ is above $0.95$ and the maximum intensity of the first side lobe reaches $0.06\%$ of the peak intensity. 
For the more realistic case of the pulse obtained after compensating the output pulse of the first stage up to the FOD, Fig. \ref{fig:Demo3}(b), the spectral broadening, phase and cleanness deteriorate as the input pulse is less intense and not as clean as in the ideal TL case. In this case, the spectrum $(\mathrm{SC}=0.54)$ is compatible with a TL pulse of $7.15$ fs FWHM duration, and in which the maximum intensity of the first side lobe reaches $0.24\%$ of the peak intensity. Regarding the spatial homogeneity, a high average V-factor is again observed, reaching up to $99.77\%$.

\subsection{Third compression stage (single-pass)}

\begin{figure}[htbp]
    \includegraphics[width=0.44\linewidth]{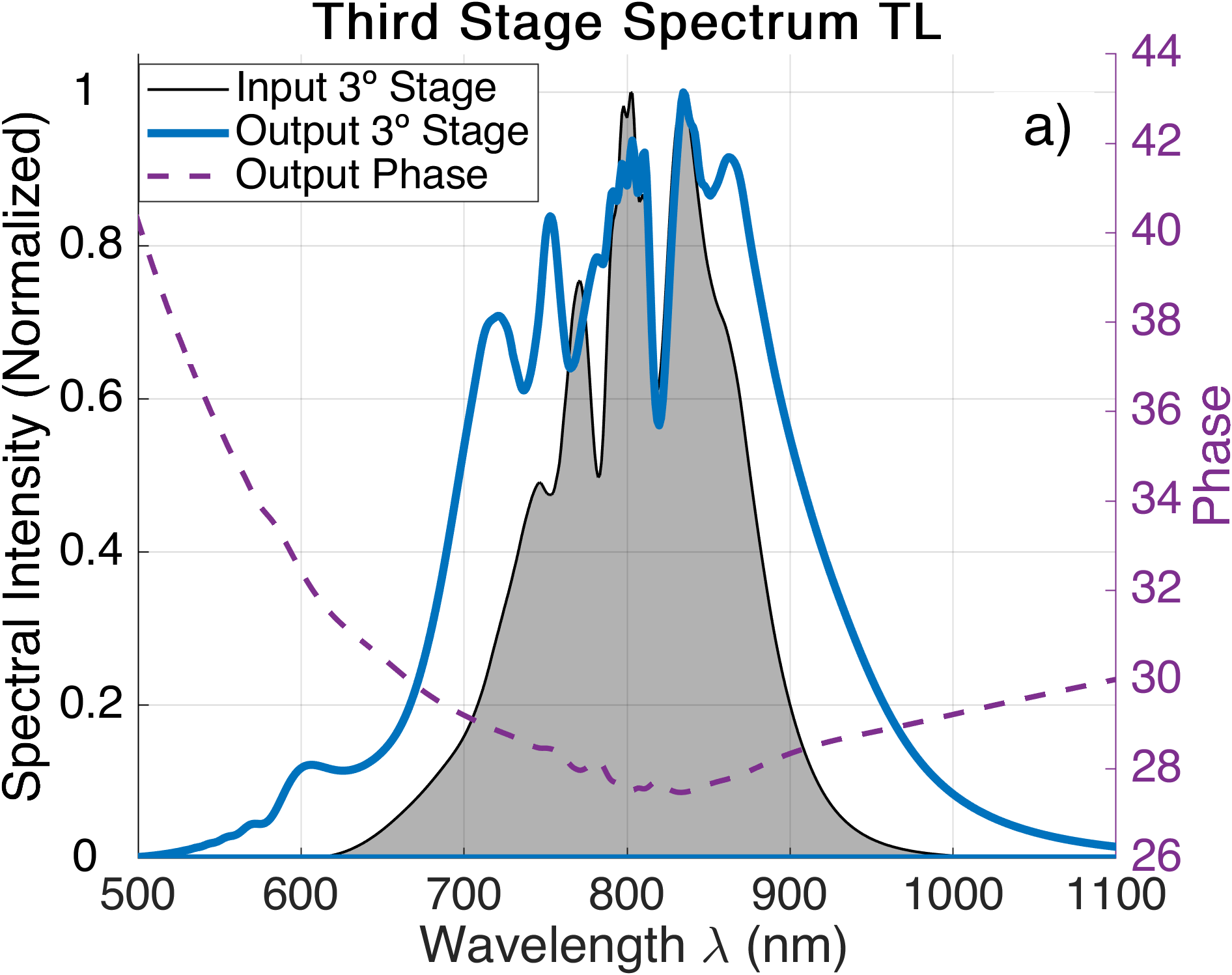}
    \includegraphics[width=0.44\linewidth]{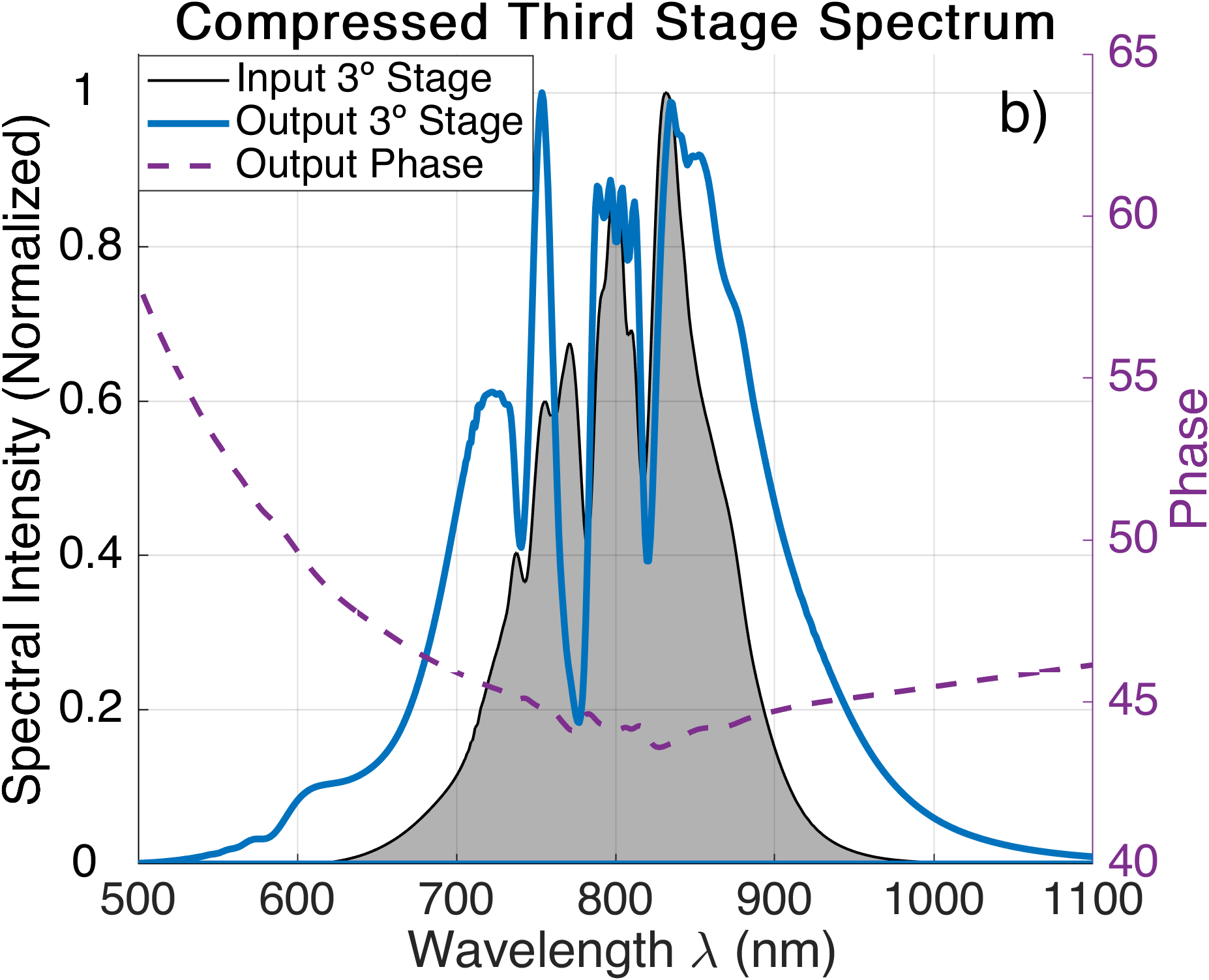}
    \centering
    \caption{On-axis output spectra and phase obtained from the third stage when following (a) the TL path or (b) a more realistic situation in which the output pulses of all stages are compensated for GDD, TOD and FOD.}
    \label{fig:Demo4}
    
\end{figure}

If we want to have a pulse in the few-cycle regime with this technique we have to repeat the process a third time. The procedure, once more, starts by compensating the output pulse obtained from the second stage and use it as the input pulse for the next stage. With the new pulse we have to design the new stage to keep in the EFCR. Taking as the input pulse for this last stage the TL pulse of the spectrum shown in Fig. \ref{fig:Demo3}(a) we obtain an input pulse with $6.73$ fs FWHM duration and a peak intensity of $1.08\times10^{13}~\mathrm{W/cm}^2$. We will call this the TL path, as we always use the TL pulse of the previous stage as the input pulse of the next one. In this case, the lengths that help us to design each stage take the following values: $L_D=0.403$ mm, $L_{NL}=0.106$ mm and $L_C=3.73$ mm. Again, the width of the plates and/or the number of round trips have to be decreased compared to the previous stage. With these typical lengths, it has not much sense to use a third MPC and it is preferable to use a single-pass thin-plate configuration.
In particular, we propose to take the beam from the compressor after the second stage and let it first propagate through vacuum as typically done in the standard thin-plate configuration \cite{zhu_spatially_2022,MingChang}. To avoid reaching the ionization threshold, we must ensure that the peak intensity remains below the previously mentioned $7.4~\mathrm{TW/cm}^2$ limit. To achieve this, we allow the beam propagate $50~\mathrm{cm}$ in free vacuum, during which the beam size increases and the peak intensity decreases to $5.96\times10^{12}~\mathrm{W/cm}^2$, resulting in an increased nonlinear length of $L_{NL}=0.193$ mm. Consequently, we must select the plate thickness to fulfill $L_{NL}<L< L_D$ and $L< L_C/10$, considering that in this case the interaction length coincides with the length of the plate, $L_I=L$. Our choice, then, is to use a 250 $\mathrm{\mu m}$ fused silica thin-plate which keeps the nonlinear propagation in the EFCR, as shown in Fig. \ref{fig:design}(b). The on-axis spectrum obtained after the thin-plate for the ideal TL path is shown in Fig. \ref{fig:Demo4}(a), which corresponds to a TL pulse of $3.80$ fs FWHM duration and in which the maximum intensity of the first side lobe reaches $0.28\%$ of the peak intensity. If we calculate a more realistic pulse (Fig. \ref{fig:Demo4}(b)) in which we compensate the three more relevant dispersion terms (GDD, TOD and FOD), we achieve a pulse with $3.97$ fs FWHM duration and in which the maximum intensity of the first side lobe reaches $0.45\%$ of the peak intensity. Either of the two results are few-cycle clean pulses as we desire. In terms of spatial performance, while a slight decrease is observed in this third stage, the homogeneity of the phase-compensated pulse remains excellent, with an average V-factor of $98.90\%$, consistent with the overall high-quality performance seen in the previous stages.

\section{Conclusion}
In this work, we have presented a comprehensive study of pulse compression using a three-stage all-bulk hybrid multipass cell (MPC) scheme, demonstrating its ability to achieve ultrashort and high-quality optical pulses.
Operating in the enhanced frequency chirp regime (EFCR), we have compressed pulses at 800 nm from 177 fs FWHM to pulses with 3.88 fs transform limited duration and side lobes whose maximum intensity value is below 0.2\% of the main peak intensity. The optimal conditions for EFCR propagation have been carefully established, enabling a spectral broadening process that is both controllable and tailored to generate spectra compatible with clean and ultrashort pulses after phase compensation. The first two stages use bulk MPCs to achieve smooth spectral broadening, while a third stage uses a thin plate to extend the spectral bandwidth while preserving the temporal quality of the pulse. After each stage, dispersion compensation is applied to maintain few-cycle pulse durations. This  all-bulk, multi-stage approach demonstrates a high degree of robustness, as the same design principles are successfully applied across different input conditions to consistently obtain the desired output. These results underscore the potential of this strategy for generating clean, few-cycle pulses suitable for applications in attosecond science and ultrafast optics.

\printbibliography
\end{document}